\documentclass[fleqn,twoside]{article}
\usepackage[headings]{espcrc2}
\usepackage{graphicx,amsmath,amssymb,axodraw2}

\newcommand\cuba{\textsc{Cuba}}
\newcommand\Vol{\mathop{\mathrm{Vol}}}
\newcommand\op[1]{\mathbf{#1}}
\newcommand\opI{\op{I}}
\newcommand\opC{\op{C}}
\newcommand\opQ{\op{Q}}
\newcommand\opM{\op{M}}
\newcommand\opN{\op{N}}
\newcommand\opL{\op{L}}
\newcommand\rd{\mathrm{d}}
\newcommand\re{\mathrm{e}}
\newcommand\ri{\mathrm{i}}
\newcommand\eg{e.g.\ }
\newcommand\ie{i.e.\ }
\newcommand\Itot{I_{\mathrm{tot}}}
\newcommand\Etot{E_{\mathrm{tot}}}
\newcommand\epsrel{\varepsilon_{\mathrm{rel}}}
\newcommand\epsabs{\varepsilon_{\mathrm{abs}}}
\newcommand\ord{\mathcal{O}}

\title{The \cuba\ Library}

\author{T. Hahn \\
	Max-Planck-Institut f\"ur Physik \\
	F\"ohringer Ring 6, D--80805 Munich, Germany}
       
\begin{document}

\begin{abstract}
Concepts and implementation of the \cuba\ library for multidimensional 
numerical integration are elucidated.
\end{abstract}

\maketitle


\section{Overview}

\cuba\ \cite{Cuba} is a library for multidimensional numerical
integration.  It features four integration algorithms with interfaces
for Fortran, C/C++, and Mathematica.  All four can integrate vector
integrands and their invocation is quite similar to make it easy to
switch routines for comparison.  The main characteristics are summarized
below.

\smallskip
\hrule
\smallskip

\noindent\emph{Routine:}
	\textbf{Vegas} \\
\emph{Basic integration methods available:} \\
	$\bullet$ Sobol sample (quasi Monte Carlo) \\
	$\bullet$ Mersenne Twister sample (pseudo Monte Carlo) \\
\emph{Variance reduction:}
	importance sampling

\smallskip
\hrule
\smallskip

\noindent\emph{Routine:}
	\textbf{Suave} \\
\emph{Basic integration methods available:} \\
	$\bullet$ Sobol sample (quasi Monte Carlo) \\
	$\bullet$ Mersenne Twister sample (pseudo Monte Carlo) \\
\emph{Variance reduction:}
	importance sampling combined with globally adaptive subdivision

\smallskip
\hrule
\smallskip

\noindent\emph{Routine:}
	\textbf{Divonne} \\
\emph{Basic integration methods available:} \\
	$\bullet$ Korobov sample (lattice method) \\
	$\bullet$ Sobol sample (quasi Monte Carlo) \\
	$\bullet$ Mersenne Twister sample (pseudo Monte Carlo) \\
	$\bullet$ cubature rules (deterministic method) \\
\emph{Variance reduction:}
	stratified sampling, aided by methods from
	numerical optimization

\smallskip
\hrule
\smallskip

\noindent\emph{Routine:}
	\textbf{Cuhre} \\
\emph{Basic integration method available:} \\
	$\bullet$ cubature rules (deterministic method) \\
\emph{Variance reduction:}
	globally adaptive subdivision

\smallskip
\hrule
\smallskip

Before explaining the buzzwords appearing in this list, it is perhaps 
worthwhile to address two frequently asked questions.

First, numerical integration rapidly becomes more difficult with
increasing dimension, no matter how many tricks are built into the
integrator.  To illustrate this, imagine computing the volume of the
$d$-dim.\ sphere $S_d$ by integrating its characteristic function $\chi
= \theta(1 - \|x\|_2)$ inside the surrounding hypercube $C_d = [-1,
1]^d$.  In a Monte Carlo way of thinking, then, the ratio $r = \Vol
S_d/\Vol C_d$ can be taken as the chance that a general-purpose
integrator will find the sphere at all.  The numbers clearly display
what is often termed the `curse of dimensionality':
\begin{equation}
\begin{array}{c||c|c|c|c}
d & 2 & 5 & 10 & 50 \\ \hline
r & .785 & .164 & .0025 & 1.5\times 10^{-28}
\end{array}
\end{equation}

Second, \cuba\ (and, for that matter, most multidimensional 
integrators) can do only Riemann integrals of the form
\begin{equation}
\opI f := \int_0^1 \rd^d x\,f(\vec x)\,.
\end{equation}
Most questions concern the boundaries, although it is straightforward to
transform a finite integration region to the unit hypercube:
\begin{gather}
\int_{a_1}^{b_1}\!\!\!\!\cdots\int_{a_d}^{b_d}\rd^d x\,f(\vec x)
= \int_0^1\rd^d y\,J\,f(\vec x) \,, \\
J = \prod_{i = 1}^d (b_i - a_i)\,,
\quad
x_i = a_i + (b_i - a_i) y_i\,. \notag
\end{gather}


\section{Concepts}

\subsection{Deterministic vs.\ Monte Carlo}

\cuba\ contains both deterministic and Monte Carlo integration methods.  
The deterministic approach is based on \emph{cubature rules},
\begin{equation}
\opI f\approx
\opC_n f := \sum_{i = 1}^n w_i f(\vec x_i)
\end{equation}
with specially chosen \emph{nodes} $\vec x_i$ and \emph{weights} $w_i$.
Error estimation is done \eg by null rules $\opN_m$ ($m < n$) which are 
constructed to give zero for functions integrated exactly by $\opC_n$ 
and thus measure errors due to ``higher terms.''

The Monte Carlo estimate, although quite similar in form,
\begin{equation}
\opI f\approx \opM_n f := \frac 1n\sum_{i = 1}^n f(\vec x_i)\,,
\end{equation}
is conceptually very different as this formula denotes the
\emph{statistical average} over independent and identically distributed
random samples $\vec x_i$.  In this case the standard deviation 
furnishes a probabilistic estimate of the integration error:
\begin{equation}
\sigma(\opM_n f) = \sqrt{\opM_n f^2 - \opM_n^2 f}\,.
\end{equation}


\subsection{Construction of Cubature Rules}

Starting from an orthogonal basis of functions $\{b_1, \dots, b_m\}$ -- 
usually monomials -- with which most $f$ can (hopefully) be approximated
sufficiently well one imposes that each $b_i$ be integrated exactly by
$\opC_n$: $\opI\,b_i \,\smash{\stackrel{\lower 3pt\hbox{\tiny !}}
{=}}\, \opC_n b_i$.  These are $m$ moment equations
\begin{equation}
\sum_{k = 1}^n w_k b_i(\vec x_k) = \int_0^1\rd^d x\,b_i(\vec x)
\end{equation}
for $n d + n$ unknowns $\vec x_i$ and $w_i$.  They pose a formidable, 
in general nonlinear, system of equations.  Additional assumptions, \eg 
symmetries, are usually necessary to solve this system.  \cuba\ employs 
the Genz--Malik rules \cite{GenzMalik} constructed from a symmetric 
monomial basis.


\subsection{Globally Adaptive Subdivision}

Once an error estimate for the integral is available, global 
adaptiveness is easy to implement:

\begin{enumerate}
\item Integrate the entire region: $\Itot\pm\Etot$.	\\[-4ex]
\item while $\Etot > \max(\epsrel\Itot, \epsabs)$	\\[-4ex]
\item\quad Find the region $r$ with the largest error.	\\[-4ex]
\item\quad Bisect (or otherwise cut up) $r$.		\\[-4ex]
\item\quad Integrate each subregion of $r$ separately.	\\[-4ex]
\item\quad $\Itot = \sum I_i$,~~
           $\Etot = \sqrt{\sum E_i^2}$.			\\[-4ex]
\item end while
\end{enumerate}

A remark is in order here about the two precisions, $\epsrel$ and
$\epsabs$.  Naively what one imposes is the relative precision: the
result is supposed to be accurate to, say, one part in a thousand, \ie
$\epsrel = 10^{-3}$.  For integral values approaching zero, however,
this goal becomes harder and harder to reach, and so as not to spend
inordinate amounts of time in such cases, an absolute precision
$\epsabs$ can be prescribed, where typically $\epsabs\ll\epsrel$.


\subsection{Importance Sampling}

Importance sampling introduces a weight function into the integral:
\begin{gather}
\opI f = \int_0^1\rd^d x\,
  w(\vec x)\,\frac{f(\vec x)}{w(\vec x)}\,, \\
w(\vec x) > 0\,,
\quad
\opI\,w = 1\,, \notag
\end{gather}
with two requirements:
a) one must be able to sample from the distribution $w(\vec x)$,
b) $f/w$ should be ``smooth'' in the sense that
   $\sigma_w(f/w) < \sigma(f)$, \eg $w$ and $f$ should have
   the same peak structure.
The ideal choice is known to be $w(\vec x) = |f(\vec x)|/\opI f$ which
has $\sigma_w(f/w) = 0$, but is of little use as it requires a-priori
knowledge of the integral value.


\subsection{Stratified Sampling}

Stratified sampling works by sampling subregions.  Consider a total of
$n$ points sampled in a region $r = r_a + r_b$ vs.\ $n/2$ points sampled
in $r_a$ and $n/2$ in $r_b$.  In the latter case the variance is
\begin{equation}
\frac 14\left(\frac{\sigma_a^2 f}{n/2} +
              \frac{\sigma_b^2 f}{n/2}\right)
= \frac{\sigma_a^2 f + \sigma_b^2 f}{2n}
\end{equation}
whereas in the former case it can be written as
\begin{equation}
\frac{\sigma^2 f}{n} =
  \frac{\sigma_a^2 f + \sigma_b^2 f}{2n} +
  \frac{(\opI_a f - \opI_b f)^2}{4n}\,.
\end{equation}
Even in this simple example the latter variance is at best equal to the
former one, and only if the integral values are identical.  The optimal
reduction of variance can be shown to occur for $n_a/n_b = \sigma_a
f/\sigma_b f$ \cite{NumRecipes}.  The recipe is thus to split up the
integration region into parts with equal variance, and then sample all
parts with same number of points.


\subsection{Quasi-Monte Carlo Methods}

Quasi-Monte Carlo methods are based on the Koksma--Hlawka inequality
which states an upper bound on the error of an integration formula
$\opQ_n f = \frac 1n\sum_{i = 1}^n f(\vec x_i)$,
\begin{equation}
|\opQ_n f - \opI f|\leqslant V(f)\,D^*(\vec x_1, \dots, \vec x_n)\,.
\end{equation}
Apart from choosing a different integrand there is little one can do 
about $V(f)$, the ``variation in the sense of Hardy and Krause.''
The \emph{discrepancy} $D^*$ of a sequence $\vec x_1, \dots, \vec x_n$
is defined as
\begin{equation}
D^* = \sup_{r\,\in\,[0, 1]^d}
  \left|\frac{\nu(r)}{n} - \Vol r\right|,
\end{equation}
where $\nu(r)$ counts the $\vec x_i$ that fall into $r$.  The word 
``equidistributed'' indeed commonly means that $\nu(r)$ is proportional 
to $\Vol r$.  Quasi-random sequences can be constructed with a 
substantially lower discrepancy than (pseudo-)random numbers.  A Monte 
Carlo algorithm based on these sequences typically achieves convergence 
rates of $\ord(\log^{d - 1} n/n)$ rather than the usual $\ord(1/\sqrt 
n)$.

\begin{figure}
\begin{center}
Sobol Quasi-Random Numbers \\
\includegraphics[width=.75\hsize]{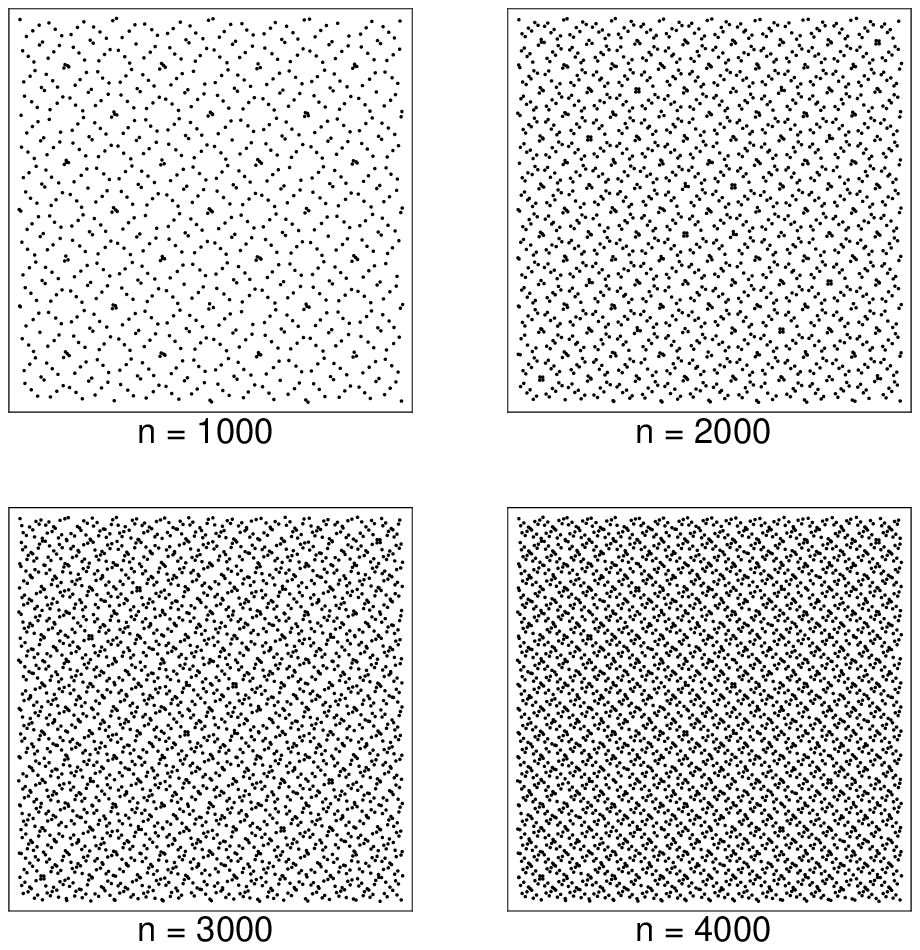} \\
Mersenne Twister Pseudo-Random Numbers \\
\includegraphics[width=.75\hsize]{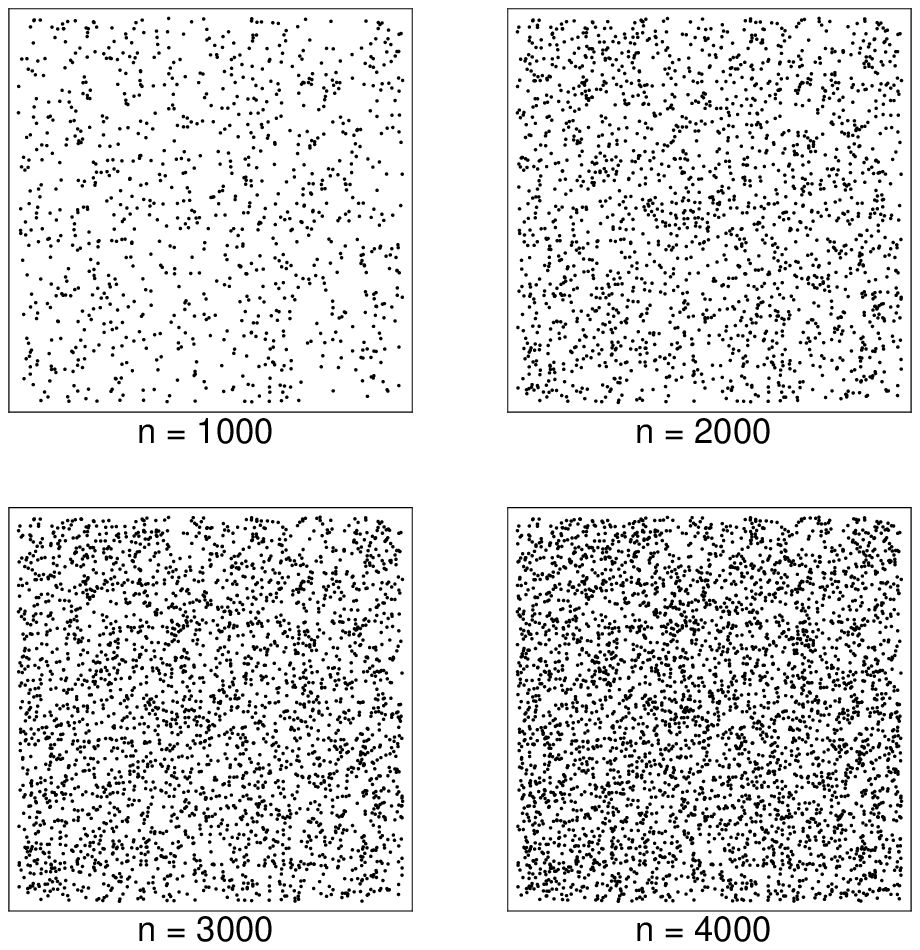}
\end{center}
\vspace*{-7ex}
\caption{\label{fig:quasipseudo}Comparison of sequences.}
\end{figure}

\cuba\ offers a choice of quasi-random Sobol sequences \cite{Sobol} or
pseudo-random Mersenne Twister sequences \cite{MersenneTwister} for all
Monte Carlo algorithms. Figure \ref{fig:quasipseudo} shows that
quasi-random numbers cover the plane much more homogeneously than
pseudo-random numbers.

\subsection{Lattice Methods}

Lattice methods require a periodic integrand, usually obtained by
applying a \emph{periodizing transformation} (\cuba's Divonne uses
$x\to |2x - 1|$).  Sampling is done on an \emph{integration lattice $L$}
spanned by a carefully selected integer vector $\vec z$:
\begin{gather}
\opL_n f = \frac 1n\sum_{i = 0}^{n - 1}
  f\bigl(\{\tfrac in\vec z\,\}\bigr)\,, \\
\{x\} = \text{fractional part of }x\,. \notag
\end{gather}
$\vec z$ is chosen (by extensive computer searches) to knock out as 
many low-order ``Bragg reflections'' as possible in the error term 
(see \eg \cite{Lattice}):
\begin{align}
\opL_n f - \opI f
&= \sum_{\vec k\in\mathbb{Z}^d}
   \tilde f(\vec k)\,\opL_n\re^{2\pi\ri\,\vec k\cdot\vec x} -
   \tilde f(\vec 0) \notag \\
&= \sum_{\vec k\in L^\perp,\,\vec k\neq\vec 0} \tilde f(\vec k)\,,
\end{align}
where $L^\perp = \{\vec k\in\mathbb{Z}^d: \vec k\cdot\vec z =
0\pmod{n}\}$ is the reciprocal lattice.


\section{Implementation}

\subsection{Vegas}

Vegas is Lepage's classic Monte Carlo algorithm \cite{Vegas}.  It uses
importance sampling for variance reduction for which it iteratively
builds up a piecewise constant weight function, represented on a
rectangular grid.  Each iteration consists of a sampling step followed
by a refinement of the grid.

In \cuba's implementation Vegas can memorize its grid for subsequent
invocations and it can save its internal state intermittently such that
the calculation can be resumed \eg after a crash.


\subsection{Suave}

Suave is a cross-breed of Vegas and Miser \cite{Miser}, a Monte Carlo
algorithm which combines Vegas-style importance sampling with globally
adaptive subdivision.

The algorithm works as follows: Until the requested accuracy is reached,
bisect the region with the largest error along the axis in which the
fluctuations of the integrand are reduced most.  Prorate the number of
new samples in each half for its fluctuation.

The Vegas grid is kept across divisions, \ie a region which is the
result of $n - 1$ subdivisions has had $n$ Vegas iterations performed on
it.  On the downside, Suave is somewhat memory hungry, as it needs to 
retain samples for later use.


\subsection{Divonne}

Divonne is a significantly extended version of CERNlib's Algorithm D151
\cite{Divonne}.  It is essentially a Monte Carlo algorithm but has
cubature rules built in for comparison, too.  Variance reduction is by
stratified sampling, which is aided by methods from numerical
optimization.  Divonne has a three-phase algorithm:

\smallskip\noindent \emph{Phase 1: Partitioning}

The integration region is split into subregions of (approximately) 
equal spread $s$, defined as
\begin{equation}
s(r) = \frac{\Vol r}{2}
  \Bigl(\max_{\vec x\in r} f(\vec x) -
        \min_{\vec x\in r} f(\vec x)\Bigr).
\end{equation}
Minimum and maximum of each subregion are sought using methods from 
numerical optimization (essentially a quasi-Newton search).  Then,
`dividers' are moved around (see picture) to find the optimal 
splitting.  This latter procedure can cleverly be translated into the 
solution of a linear system and is hence quite fast (for details see 
\cite{Divonne}).
\begin{center}
\begin{picture}(70,70)(0,0)
\SetScale{.7}
\CBox(5,5)(95,95){Blue}{PastelBlue}
\SetWidth{1}
\SetColor{Red}
\Line(0,20)(100,20)
\Line(0,65)(100,65)
\Line(30,0)(30,100)
\Line(85,0)(85,100)
\SetWidth{0}
\SetColor{Black}
\LongArrow(1,18)(1,14)
\LongArrow(1,22)(1,26)
\LongArrow(1,63)(1,59)
\LongArrow(1,67)(1,71)
\LongArrow(28,1)(24,1)
\LongArrow(32,1)(36,1)
\LongArrow(83,1)(79,1)
\LongArrow(87,1)(91,1)
\end{picture}
\end{center}

\smallskip\noindent \emph{Phase 2: Sampling}

The subregions determined in Phase 1 are independently sampled with the 
same number of points each.  The latter is extrapolated from the results 
of Phase 1.

\smallskip\noindent \emph{Phase 3: Refinement}

Regions whose results from Phase 1 and 2 do not agree within their
errors are subdivided or sampled again.  This phase is an addition to
the original algorithm since it was found that often enough the error
estimate, or even the integral value, was off because characteristics of
the integrand had not been found in Phase 1.

Two important features have been added in the \cuba\ implementation:
\begin{itemize}
\item The user can point out extrema for tricky integrands.

\item For integrands which cannot be sampled too close to the border,
      a `safety distance' can be prescribed within which values will be
      extrapolated from two points in the interior.
\end{itemize}


\subsection{Cuhre}

Cuhre is a deterministic algorithm.  It uses the Genz--Malik cubature
rules \cite{GenzMalik} in a globally adaptive subdivision scheme.  The
algorithm is thus: Until the requested accuracy is reached, bisect the
region with the largest error along the axis with the largest fourth
difference.

Cuhre has been re-implemented in \cuba\ mostly for a consistent
interface, it is the same as the original DCUHRE subroutine
\cite{dcuhre}.


\section{Comparison}

Doing a balanced comparison on integration algorithms is nearly
hopeless.  Performance depends highly on the integrand and there are
always cases, and not just academic ones, in which one routine
outperforms the others, or conversely, in which one routine simply gives
wrong results.  This, of course, is the main reason why there are four
independent and easily interchangeable algorithms in the \cuba\ library.

In this context it should be pointed out that the statistical error
estimate quoted by the Monte Carlo algorithms merely states the
one-sigma interval, or in other words: the probability that the central
value lies in the given interval is (only) about 68\%.

With these caveats, the following plot compares the performance of the 
four \cuba\ routines on a real phase-space integration.  The results 
obtained by the four methods (not shown here) indeed agree to within the 
requested accuracy.
\begin{center}
\includegraphics[width=\linewidth]{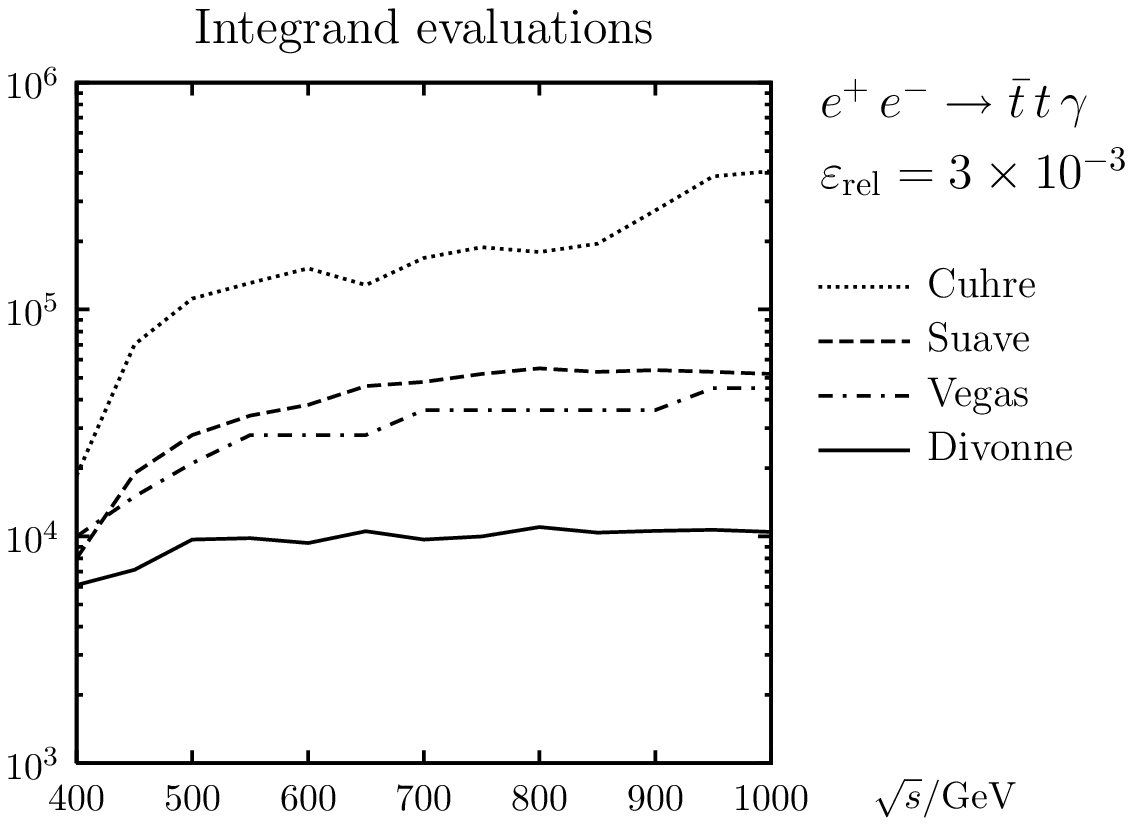}
\end{center}


\section{Mathematica interface}

The Mathematica interface deserves a special mention, as it is not a
library in the proper sense.  It is rather four executables which
communicate with Mathematica via the MathLink API:

\begin{center}
\begin{picture}(210,70)
\SetScale{.75}
\Text(50,85)[b]{Mathematica}
\CBox(0,0)(100,80){Red}{PastelRed}
\Text(50,65)[]{\Black{\small\texttt{Vegas[$f$, $\dots$]}}}
\Text(50,30)[]{\Black{\small integrand $f$}}
\Text(50,15)[]{\Black{\scriptsize (compiled function)}}
\SetOffset(135,0)
\Text(50,85)[b]{C}
\CBox(0,0)(100,80){Blue}{PastelBlue}
\Text(50,65)[]{\Black{\small\texttt{void Vegas($\dots$)}}}
\Text(50,25)[]{\Black{\small request samples}}

\SetOffset(105,0)
\SetWidth{1.5}
\LongArrow(-36,65)(36,65)
\Text(0,71)[b]{\small MathLink}

\Text(0,34)[b]{\small $\{\vec x_1, \vec x_2, \dots\}$}
\LongArrow(36,30)(-36,30)
\LongArrow(-36,20)(36,20)
\Text(0,16)[t]{\small $\{f_1, f_2, \dots\}$}
\end{picture}
\end{center}
After loading the appropriate MathLink executable, \eg with
\texttt{Install["Vegas"]}, the corresponding routine can be used almost
like Mathematica's native \texttt{NIntegrate}.  The integrand is
evaluated completely in Mathematica, which means that one can do things
like
\begin{verbatim}
   Cuhre[Zeta[x y], {x,2,3}, {y,4,5}]
\end{verbatim}


\section{Further Tools}

\subsection{Chooser}

\cuba\ includes a ``one-stop interface'' which further simplifies the
invocation:
\begin{verbatim}
  subroutine Cuba(method, ndim, ncomp,
    integrand, integral, error, prob)
\end{verbatim}
The user just has to choose \texttt{method = 1,2,3,4} to switch between
Vegas, Suave, Divonne, Cuhre.  All parameters specific to individual 
routines are ``hidden'' (determined inside the routine), \ie this is not 
a finished product, but should be adapted by the user.


\subsection{Partition Viewer}

\cuba's Partition Viewer displays the partition taken by the integration
algorithm.  This is sometimes helpful to visualize where the integrand's
characteristic regions lie.  It is really useful only in small to
moderate dimensions, though.

\begin{figure}
\centerline{\includegraphics[height=.9\linewidth]{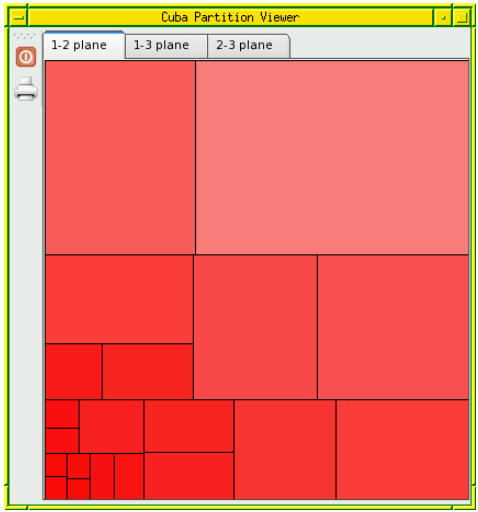}}
\vspace*{-3ex}
\caption{\label{fig:partview}Screenshot of \cuba's partition viewer.}
\end{figure}

Verbosity level 3 must be chosen in the integration routine and the 
output piped through the \texttt{partview} utility, as in
\begin{verbatim}
  myprogram | partview 1 2
\end{verbatim}
which will then display the 1--2 plane of the partitioning.
Figure \ref{fig:partview} shows a screenshot.

\section{Summary}

\cuba\ is a library for multidimensional numerical integration written
in C.  It contains four independent algorithms: Vegas, Suave, Divonne,
and Cuhre which have similar invocations and can be exchanged easily for
comparison.  All routines can integrate vector integrands and have a
Fortran, C/C++, and Mathematica interface.  Additional tools are
included, such as a one-stop invocation and a partition viewer.  \cuba\ 
is available at \texttt{http://www.feynarts.de/cuba}, licensed under the 
LGPL, and easy to build (autoconf).


\end{document}